# Sparse Channel Estimation for MIMO-OFDM Amplify-and-Forward Two-Way Relay Networks

Guan Gui, Wei Peng and Fumiyuki Adachi
Department of Communications Engineering,
Graduate School of Engineering, Tohoku University
6-6-05 Aza-Aoba, Aramaki, Aoba-ku, Sendai, 980-8579 Japan
E-mail: {gui,peng}@mobile.ecei.tohoku.ac.jp, adachi@ ecei.tohoku.ac.jp

*Abstract*—Accurate channel impulse response (CIR) is required for coherent detection and it can also help improve communication quality of service in next-generation wireless communication systems. One of the advanced systems is multi-input multi-output orthogonal frequency-division multiplexing (MIMO-OFDM) amplify and forward two-way relay networks (AF-TWRN). Linear channel estimation methods, e.g., least square (LS), have been proposed to estimate the CIR. However, these methods never take advantage of channel sparsity and then cause performance loss. In this paper, we propose a sparse channel estimation method to exploit the sparse structure information in the CIR at each end user. Sparse channel estimation problem is formulated as compressed sensing (CS) using sparse decomposition theory and the estimation process is implemented by LASSO algorithm. Computer simulation results are given to confirm the superiority of proposed method over the LS-based channel estimation method.

I. INTRODUCTION

It is well known that wireless communication technologies are developing rapidly due to that the huge market is promoted by the skyrocketing number of wireless users is in last decades [1][1]. Until now, there have three promising techniques for broadband wireless communications. The first technique is multiple antenna transmission over multi-input multi-output (MIMO) that is becoming one of the prevail techniques for enhancing system capacity and combating multipath channel fading. The second technique is orthogonal frequency division multiplexing (OFDM) modulation which provides high spectral efficient and robustness mitigates frequency-selective channel fading [1]. The third technique is two-way relay network (TWRN) that implements information exchange in two time slots. When comparing with four-time slots traditional TWRN (see Fig. 1(a)) and three time slots physical-layer TWRN (see Fig. 1(b)) which achieve information exchange, two time slots TWRN (see Fig. 3(c)) can enhances system capacity 66.7% and 100%, respectively. In addition, TWRN can also improve transmission range with limited transmitted power [2]. To take advantage of three techniques fully, combine them into an advanced wireless communication system is one of promising candidate techniques. However, one of the key challenges is how to obtain accurate channel state information (CSI) which is applied for self-interference removal and coherent detection at each terminal.

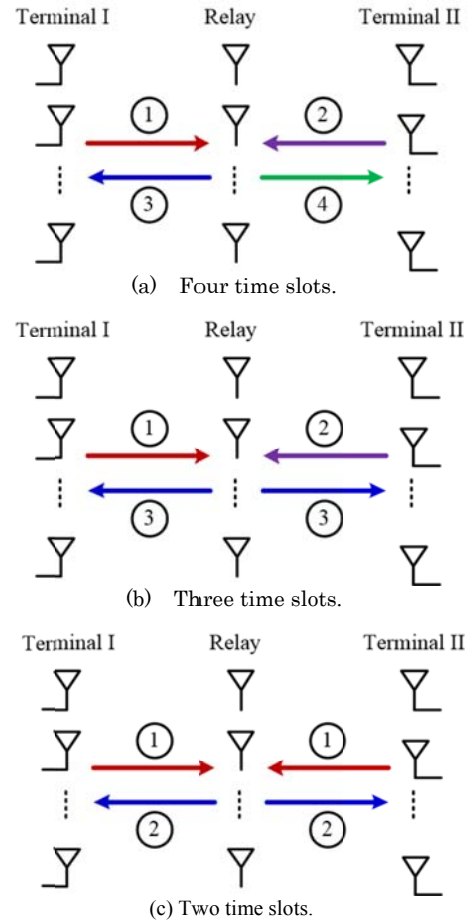

Fig. 1. Information exchange using different time slots in MIMO-OFDM AF-TWRN.

In this paper, channel estimation problem will be studied in a cyclic prefix (CP) based MIMO-OFDM TWRN. Here, decode-and-forward (DF) based TWRN system should be done channel estimation at both terminals and relay station for coherent detection. Channel estimation and signal modulation will increase high computation burden on the relay. In addition, channel estimation techniques in DF MIMO-OFDM-TWRN can be borrowed from point-to-point MIMO-OFDM systems [3–7]. Unlike the DF model, the obvious advantage of AF one can alleviate the computational burden on the relay, i.e., the relay amplifies and forwards the signals received from both terminals. Due to the signaling rule, hence, only the cascaded channels are necessary for self-interference removal and coherent detection at each terminal. Except that the advantage of low computational burden at the relay, estimate cascaded channels at each terminal can mitigate the quantization error and also can avoid individual channels further distortion from noise [8].

Traditional linear channel estimation methods, e.g, LS [9], have been proposed for MIMO-OFDM AF-TWRN . However, these methods cannot take the advantage of inherent channel sparsity and hence cause performance loss. In this paper, we propose sparse channel approach to exploit such channel sparsity. Sparse channel estimation problem in MIMO-OFDM AF-TWRN is formulated as CS problem. At each terminal, equivalent training signal is constructed to probe equivalent channel vector using least absolute shrinkage and selection operator (LASSO) [10]. The performance of propose method will be evaluated by computer simulations.

The remainder of this paper is organized as follows. A MIMO-OFDM AF-TWRN system model is described and problem formulation is given in Section II. In section III, the sparse channel estimation method is proposed and lower bound of estimation performance is derived. Computer simulation results are given in Section IV in order to evaluate and compare performances of LS-based channel estimation method. Finally, we conclude the paper in Section V.

## II. SYSTEM MODEL AND PROBLEM FORMUALTION

As shown in Fig. 1(c), we consider a MIMO-OFDM AF-TWRN in which two-time slots information exchange between terminal $\mathbf{T}_1$ and terminal $\mathbf{T}_2$ with the help of relay $\mathbf{R}$. Both the two terminals and the relay have $N_t$ and $N_r$ antennas ($N_r \geq N_t$), respectively. Assume that $L$-length channel vectors between the $n_t$-th antenna of terminals $\mathbf{T}_i$, $i = 1,2$ and $n_r$-th antenna of relay $\mathbf{R}$ are denoted by $\mathbf{h}_{n_t n_r} = [h_{n_t n_r}(0), h_{n_t n_r}(1), ..., h_{n_t n_r}(L-1)]^T$ and $\boldsymbol{g}_{n_t n_r} = [g_{n_t n_r}(0), g_{n_t n_r}(1), ..., g_{n_t n_r}(L-1)]^T$, respectively. Each channel vector is supported only by $K$ nonzero taps and $K \ll L$. Suppose that each the nonzero tap is modeled as a complex Gaussian random variable with zero mean and variance $\sigma_{h,l}^2$, and $\sigma_{g,l}^2$ $l = 0,1,...,L$. In addition, $\mathbf{h}_{n_t n_r}$ and $\boldsymbol{g}_{n_t n_r}$ are assumed invariant in the two time slots information exchange. At time $t$, suppose that OFDM signal vectors are transmitted from $n_t$-th antenna of $\mathbf{T}_i$, $i = 1,2$, are $\bar{\mathbf{s}}_{n_t} = [\bar{s}_{n_t}(0), \bar{s}_{n_t}(1), ..., \bar{s}_{n_t}(N-1)]^T$ and $\bar{\mathbf{x}}_{n_t} = [\bar{x}_{n_t}(0), \bar{x}_{n_t}(1), ..., \bar{x}_{n_t}(N-1)]^T$, respectively, where $N$ is the number of subcarriers and $n_r = 1,2,..,N_r$. At the same time, two transmitted power is assumed $E[\bar{\mathbf{s}}_{n_t}^H \bar{\mathbf{s}}_{n_t}] = NP_1$ and $E[\bar{\mathbf{x}}_{n_t}^H \bar{\mathbf{x}}_{n_t}] = NP_1$, respectively.

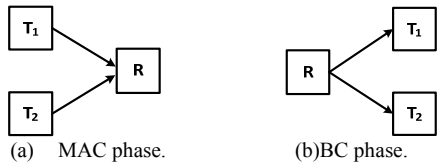

(a) MAC phase. (b) BC phase.

Fig. 2. Information exchanges under TWNR.

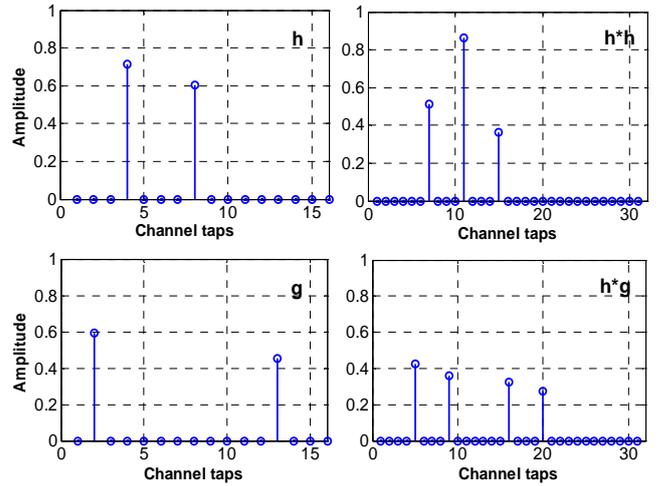

Fig. 3. Example of two individual channels and their cascaded ones.

### A. MAC phase

In the multi-access (MAC) phase as shown in Fig. 2(a), inverse discrete Fourier transform (IDFT) is computed for frequency-domain signal vectors $\bar{\mathbf{s}}_{n_t}$ and $\bar{\mathbf{x}}_{n_t}$. The resultant vectors, $\mathbf{s}_{n_t} = \mathbf{F}^H \bar{\mathbf{s}}_{n_t}$ and $\mathbf{x}_{n_t} = \mathbf{F}^H \bar{\mathbf{x}}_{n_t}$, are then cyclic prefix (CP) padded with length $L_{CP} \geq (L-1)$ to avoid inter-block interference (IBI). Here, $\mathbf{F}$ is a $N \times N$ discrete Fourier transform (DFT) matrix where entries $f_{mn} = 1/N\, e^{-j\pi mn/N}$, $m, n = 0,1,...,N$. After removed the CP, the received signal vector at the $n_r$-th antenna of $\mathbf{R}$ for $t = 1,2,...,T$ is written as

$$\mathbf{r}_{n_r} = \sum_{n_t}^{N_t} \mathbf{H}_{n_t n_r} \mathbf{s}_{n_t} + \sum_{n_t}^{N_t} \mathbf{G}_{n_t n_r} \mathbf{x}_{n_t} + \mathbf{z}_{n_r}, \quad (1)$$

for $n_r = 1,2,...,N_r$, where $\mathbf{H}_{n_t n_r}$ and $\mathbf{G}_{n_t n_r}$ are circulant matrices with the first columns of $[\mathbf{h}_{n_t n_r}^T, \mathbf{0}_{1 \times (N-L)}]^T$, and $[\mathbf{g}_{n_t n_r}^T, \mathbf{0}_{1 \times (N-L)}]^T$, respectively. The additive noise vector $\mathbf{z}_{n_r}$ satisfies $CN(\mathbf{0}_{N \times 1}, \sigma_n^2 \mathbf{I}_N)$. If we collect all received signal vectors $\mathbf{r}_{n_r}$ $n_r = 1,2,...,N_r$ at $\mathbf{R}$ to form a $N_r N$-length vector $\mathbf{r} = [\mathbf{r}_1^T, ..., \mathbf{r}_{n_r}^T, ..., \mathbf{r}_{N_r}^T]^T$, then the received model in the MAC phase at relay is written as

$$\mathbf{r} = \mathbf{Hs} + \mathbf{Gx} + \mathbf{z}, \quad (2)$$

where

$$\mathbf{H} = \begin{bmatrix} \mathbf{H}_{11} & \mathbf{H}_{21} & \cdots & \mathbf{H}_{N_t 1} \\ \mathbf{H}_{12} & \mathbf{H}_{22} & \cdots & \mathbf{H}_{N_t 2} \\ \vdots & \vdots & \cdots & \vdots \\ \mathbf{H}_{1N_r} & \mathbf{H}_{2N_r} & \cdots & \mathbf{H}_{N_t N_r} \end{bmatrix} \in \mathbb{C}^{N_r N \times N_t N}, \quad (3)$$

$$\mathbf{G} = \begin{bmatrix} \mathbf{G}_{11} & \mathbf{G}_{21} & \cdots & \mathbf{G}_{N_t 1} \\ \mathbf{G}_{12} & \mathbf{G}_{22} & \cdots & \mathbf{G}_{N_t 2} \\ \vdots & \vdots & \cdots & \vdots \\ \mathbf{G}_{1N_r} & \mathbf{G}_{2N_r} & \cdots & \mathbf{G}_{N_t N_r} \end{bmatrix} \in \mathbb{C}^{N_r N \times N_t N}, \quad (4)$$

$$\mathbf{s} = [\mathbf{s}_1^T, ..., \mathbf{s}_{n_t}^T, ..., \mathbf{s}_{N_t}^T]^T \in \mathbb{C}^{N_t N \times 1}, \quad (5)$$

$$\mathbf{x} = [\mathbf{x}_1^T, ..., \mathbf{x}_{n_t}^T, ..., \mathbf{x}_{N_t}^T]^T \in \mathbb{C}^{N_t N \times 1}, \quad (6)$$

$$\mathbf{z} = [\mathbf{z}_1^T, ..., \mathbf{z}_{n_r}^T, ..., \mathbf{z}_{N_r}^T]^T \in \mathbb{C}^{N_r N \times 1}, \quad (7)$$

According to Eq. (2), the received signal vector $\mathbf{r}_{n_r}$ is amplified by a positive coefficients $\beta$ which is given by

$$\beta = \sqrt{\frac{P_r}{N_t \sum_{l=0}^{L-1}(\sigma_{h,l}^2 P_1 + \sigma_{g,l}^2 P_2) + N_0}}, \quad (8)$$

where $P_r$ is relay's amplify power which is given by $\mathrm{E}[\bar{\mathbf{r}}_{n_r}^H \bar{\mathbf{r}}_{n_r}] = N P_r$.

*B. BC phase*

Because of system symmetrical in TWRN, without loss of generality, we consider the broadcasting (BC) phase at $\mathbf{T}_1$, as shown in Fig.2(b). Let $\mathbf{y}_{n_t}$ denote the received signal vectors at the $n_t$-th antenna at time (t + T). If we collect $N_t$ received vectors $\mathbf{y}_{n_t}$ as $\mathbf{y} = [\mathbf{y}_1^T, \mathbf{y}_2^T, \dots, \mathbf{y}_{N_t}^T]^T$, then received signal model can be written as

$$\mathbf{y} = \beta \widetilde{\mathbf{H}} \mathbf{H} \mathbf{s} + \beta \widetilde{\mathbf{H}} \mathbf{G} \mathbf{x} + \beta \widetilde{\mathbf{H}} \mathbf{z} + \mathbf{v}, \quad (9)$$

where

$$\widetilde{\mathbf{H}} = \begin{bmatrix} \mathbf{H}_{11} & \mathbf{H}_{12} & \dots & \mathbf{H}_{1N_r} \\ \mathbf{H}_{21} & \mathbf{H}_{22} & \dots & \mathbf{H}_{2N_r} \\ \vdots & \vdots & \dots & \vdots \\ \mathbf{H}_{N_t 1} & \mathbf{H}_{N_t 2} & \dots & \mathbf{H}_{N_t N_r} \end{bmatrix} \in \mathbb{C}^{N_t N \times N_r N}, \quad (10)$$

$$\mathbf{v} = [\mathbf{v}_1^T, \mathbf{v}_2^T, \dots, \mathbf{v}_{N_t}^T]^T \in \mathbb{C}^{N_t N \times 1} \quad (11)$$

where $\mathbf{v}_{n_t}$ is a noise vector at $n_t$-th antenna of $\mathbf{T}_1$, satisfying $\mathbf{v}_{n_t} \in \mathrm{CN}(\mathbf{0}_{N \times 1}, \sigma_n^2 \mathbf{I}_N)$. According to matrix theory [11], circulant matrices $\mathbf{H}_{n_t n_r}$ and $\mathbf{G}_{n_t n_r}$, $n_r = 1,2,\dots,N_r$, $n_t = 1,2,\dots,N_t$, can be decomposed as

$$\mathbf{H}_{n_t n_r} = \mathbf{F}^H \boldsymbol{\Lambda}_{n_t n_r} \mathbf{F}, \quad (12)$$

$$\mathbf{G}_{n_t n_r} = \mathbf{F}^H \mathbf{U}_{n_t n_r} \mathbf{F}, \quad (13)$$

respectively, where $(\cdot)^H$ denotes matrix Hermitian transition operation and above diagonal matrices are given by

$$\boldsymbol{\Lambda}_{n_t n_r} = \mathrm{diag}\{H_{n_t n_r}(0), \dots, H_{n_t n_r}(n), \dots, H_{n_t n_r}(N-1)\}, \quad (14)$$

$$\mathbf{U}_{n_t n_r} = \mathrm{diag}\{G_{n_t n_r}(0), \dots, G_{n_t n_r}(n), \dots, G_{n_t n_r}(N-1)\}, \quad (15)$$

respectively. Based on the above analysis, it is easy found that the $n$-th diagonal entries $H_{n_t n_r}(n)$ in Eq. (14) and $G_{n_t n_r}(n)$ in Eq. (15) are obtained by

$$H_{n_t n_r}(n) = \sum_{l=0}^{L-1} h_{n_t n_r}(n) e^{-j2\pi nl/N}, \quad (16)$$

$$G_{n_t n_r}(n) = \sum_{l=0}^{L-1} g_{n_t n_r}(n) e^{-j2\pi nl/N}, \quad (17)$$

respectively. Therefore, the product of $\beta \mathbf{H}_{n_t n_r} \mathbf{H}_{n_t' n_r'}$ and $\beta \mathbf{G}_{n_t n_r} \mathbf{H}_{n_t' n_r'}$ with respect to $n_t, n_t' = 1,2,\dots,N_t$ and $n_r = 1,2,\dots,N_r$ can also be written as

$$\beta \widetilde{\mathbf{H}}_{n_t n_r} \mathbf{H}_{n_t' n_r} = \mathbf{F}^H \beta \boldsymbol{\Lambda}_{n_t n_r} \boldsymbol{\Lambda}_{n_t' n_r} \mathbf{F}, \quad (18)$$

$$\beta \widetilde{\mathbf{H}}_{n_t n_r} \mathbf{G}_{n_t n_r} = \mathbf{F}^H \beta \boldsymbol{\Lambda}_{n_t n_r} \mathbf{U}_{n_t n_r} \mathbf{F}, \quad (19)$$

respectively. Hence, both $\beta \widetilde{\mathbf{H}}_{n_t n_r} \mathbf{H}_{n_t' n_r}$ and $\beta \widetilde{\mathbf{H}}_{n_t n_r} \mathbf{G}_{n_t n_r}$ are circulant matrices where their first columns are given by $[\beta(\mathbf{h}_{n_t n_r} * \mathbf{h}_{n_t' n_r})^T \quad \mathbf{0}_{1 \times (N-2L+1)}]^T$ and $[\beta(\mathbf{h}_{n_t n_r} * \mathbf{g}_{n_t n_r})^T$ $\mathbf{0}_{1 \times (N-2L+1)}]^T$, respectively, where '*' denotes convolution operator between two channel vectors. Based on this observation, when the $n_1$-th row partitions of $\beta \widetilde{\mathbf{H}}$ multiplies with the $n_2$-th column partitions of $\mathbf{H}$, $n_t, n_t' = 1,2,\dots,N_t$, we can obtain an equivalent (2L-1)-length cascaded channel vector $\mathbf{q}_{n_t' n_t} \triangleq [q_{n_t' n_t}(0), \dots, q_{n_t' n_t}(l), \dots, q_{n_t' n_t}(2L-2)]^T$ which is given by

$$\mathbf{q}_{n_t' n_t} \triangleq \beta \sum_{n_r=1}^{N_r} \mathbf{h}_{n_t n_r} * \mathbf{h}_{n_t' n_r}. \quad (20)$$

Because of the symmetry of two MIMO channel matrices, we can easy find their symmetry relationship, that is, $\mathbf{q}_{n_t' n_t} = \mathbf{q}_{n_t n_t'}$. Hence, the product $\beta \widetilde{\mathbf{H}} \mathbf{H}$ is equivalent to provide $(N_t^2 + N_t)/2$ independent $(2L-1)$-length composite channel vectors $\mathbf{q}_{n_t' n_t}$ with $n_t, n_t' = 1,2,\dots,N_t$. Note that $(N_t^2 + N_t)/2 < N_t^2$ if $N_t > 1$. By virtual of the duplication matrix property [12] on sparse channel estimation, it can reduce some amount of complexity which relates to the number of antenna $N_t$, especially in the case of a relatively large scale communication system. That is to say, the computational complexity is reduce to $O((N_t^2 + N_t)/2)$ rather than $O(N_t^2)$, where $O(\cdot)$ denotes the calculation metric of complexity. Due to independent between the two MIMO channel matrices $\widetilde{\mathbf{H}}$ and $\mathbf{G}$, hence, $\beta \widetilde{\mathbf{H}} \mathbf{G}$ is equivalent to generate $N_t^2$ independent (2L-1)-length cascaded channel vectors $\mathbf{p}_{n_t n_t'} \triangleq [p_{n_t n_t'}(0), \dots, p_{n_t n_t'}(l), \dots, p_{n_t n_t'}(2L-2)]^T$ with respect to $n_t, n_t' = 1,2,\dots,N_t$, where

$$\mathbf{p}_{n_t n_t'} \triangleq \beta \sum_{n_r=1}^{N_r} \mathbf{h}_{n_t n_r} * \mathbf{g}_{n_t' n_r}. \quad (21)$$

If we define $\widetilde{\mathbf{F}} = \mathbf{I}_{N_t} \otimes \mathbf{F} \in \mathbb{C}^{N_t N \times N_r N}$, where '$\otimes$' denotes Kronecker product and $\mathbf{I}_{N_t}$ denotes an $N_t \times N_t$ identity matrix, the received signal $\mathbf{y}$ in Eq. (9) is transformed to frequency-domain using DFT matrix $\widetilde{\mathbf{F}}$, then, we have

$$\bar{\mathbf{y}} = \widetilde{\mathbf{F}} \beta \widetilde{\mathbf{H}} \mathbf{H} \widetilde{\mathbf{F}}^H \bar{\mathbf{s}} + \widetilde{\mathbf{F}} \beta \widetilde{\mathbf{H}} \mathbf{G} \widetilde{\mathbf{F}}^H \bar{\mathbf{x}} + \bar{\mathbf{v}}, \quad (22)$$

where $\bar{\mathbf{v}} = \widetilde{\mathbf{F}} \beta \widetilde{\mathbf{H}} \mathbf{z} + \widetilde{\mathbf{F}} \mathbf{v}$ denotes composite noise vector at the $\mathbf{T}_1$. According to Eq. (18) and (19), $\widetilde{\mathbf{F}} \beta \widetilde{\mathbf{H}} \mathbf{H} \widetilde{\mathbf{F}}^H$ and $\widetilde{\mathbf{F}} \beta \widetilde{\mathbf{H}} \mathbf{G} \widetilde{\mathbf{F}}^H$ can be given in Eq. (23) and (24), respectively. If we define both $\mathbf{S}_i = \mathrm{diag}(\bar{\mathbf{s}}_i)$ and $\mathbf{X}_i = \mathrm{diag}(\bar{\mathbf{x}}_i)$ as $N \times N$ diagonal matrices, and collect all cascaded channel vectors as $\mathbf{q} \triangleq [\mathbf{q}_{11}^T, \dots, \mathbf{q}_{1N_t}^T, \mathbf{q}_{22}^T, \dots, \mathbf{q}_{2N_t}^T, \dots, \mathbf{q}_{N_t^2}^T]^T$ and $\mathbf{p} \triangleq [\mathbf{p}_{11}^T, \dots, \mathbf{p}_{N_t 1}^T, \dots, \mathbf{p}_{1N_t}^T, \dots, \mathbf{p}_{N_t^2}^T]^T$, then two equivalent training signal matrices can be written in Eq. (25) and (26) respectively, where $\mathbf{F}_{2L-1}$ is partial DFT matrix by extracting the first $(2L-1)$-columns of $\mathbf{F}$.

Then the received signal model in Eq. (22) can be reformulated as

$$\bar{\mathbf{y}} = \mathbf{S}\mathbf{q} + \mathbf{X}\mathbf{p} + \bar{\mathbf{v}} = \mathbf{D}\mathbf{b} + \bar{\mathbf{v}}, \quad (1)$$

where $\mathbf{D} = [\mathbf{S}, \mathbf{X}]$ denotes an equivalent training matrix combined two training signal matrices $\mathbf{S}$ of $N_t N \times (2L-1) N_t (N_t + 1)/2$ sizes and $\mathbf{X}$ of $N_t N \times (2L-1) N_t^2$ sizes; and $\mathbf{b} = [\mathbf{q}^T, \mathbf{p}^T]^T$ denotes overall channel vector including $\mathbf{q}$ and $\mathbf{p}$. At the receive side of $\mathbf{T}_1$, channel estimator $\mathbf{q}$ is used to remove self-data interference and channel estimator $\mathbf{p}$ is applied to extract other users' data information

$$\tilde{\mathbf{F}}\beta\tilde{\mathbf{H}}\mathbf{H}\tilde{\mathbf{F}}^H = \begin{bmatrix} \sum_{n_r=1}^{N_r}\beta\Lambda_{1n_r}\Lambda_{1n_r} & \sum_{n_r=1}^{N_r}\beta\Lambda_{1n_r}\Lambda_{2n_r} & \cdots & \sum_{n_r=1}^{N_r}\beta\Lambda_{1n_r}\Lambda_{N_tn_r} \\ \sum_{n_r=1}^{N_r}\beta\Lambda_{2n_r}\Lambda_{1n_r} & \sum_{n_r=1}^{N_r}\beta\Lambda_{2n_r}\Lambda_{2n_r} & \cdots & \sum_{n_r=1}^{N_r}\beta\Lambda_{2n_r}\Lambda_{N_tn_r} \\ \vdots & \vdots & \ddots & \vdots \\ \sum_{n_r=1}^{N_r}\beta\Lambda_{N_tn_r}\Lambda_{1n_r} & \sum_{n_r=1}^{N_r}\beta\Lambda_{N_tn_r}\Lambda_{2n_r} & \cdots & \sum_{n_r=1}^{N_r}\beta\Lambda_{N_tn_r}\Lambda_{N_tn_r} \end{bmatrix} \in \mathbb{C}^{N_tN \times N_tN}, \quad (23)$$

$$\tilde{\mathbf{F}}\beta\tilde{\mathbf{H}}\mathbf{G}\tilde{\mathbf{F}}^H = \begin{bmatrix} \sum_{n_r=1}^{N_r}\beta\Lambda_{1n_r}\mathbf{U}_{1n_r} & \sum_{n_r=1}^{N_r}\beta\Lambda_{1n_r}\mathbf{U}_{2n_r} & \cdots & \sum_{n_r=1}^{N_r}\beta\Lambda_{1n_r}\mathbf{U}_{N_tn_r} \\ \sum_{n_r=1}^{N_r}\beta\Lambda_{2n_r}\mathbf{U}_{1n_r} & \sum_{n_r=1}^{N_r}\beta\Lambda_{2n_r}\mathbf{U}_{2n_r} & \cdots & \sum_{n_r=1}^{N_r}\beta\Lambda_{2n_r}\mathbf{U}_{N_tn_r} \\ \vdots & \vdots & \ddots & \vdots \\ \sum_{n_r=1}^{N_r}\beta\Lambda_{N_tn_r}\mathbf{U}_{1n_r} & \sum_{n_r=1}^{N_r}\beta\Lambda_{N_tn_r}\mathbf{U}_{2n_r} & \cdots & \sum_{n_r=1}^{N_r}\beta\Lambda_{N_tn_r}\Lambda\mathbf{U}_{N_tn_r} \end{bmatrix} \in \mathbb{C}^{N_tN \times N_tN}, \quad (24)$$

$$\mathbf{S} = \begin{bmatrix} \mathbf{S}_1\mathbf{F}_{2L-1} & \mathbf{S}_2\mathbf{F}_{2L-1} & \mathbf{S}_3\mathbf{F}_{2L-1} & \cdots & \mathbf{S}_{N_t}\mathbf{F}_{2L-1} & \mathbf{0}_{N\times(2L-1)} & \mathbf{0}_{N\times(2L-1)} & \mathbf{0}_{N\times(2L-1)} \\ \mathbf{0}_{N\times(2L-1)} & \mathbf{S}_1\mathbf{F}_{2L-1} & \mathbf{S}_2\mathbf{F}_{2L-1} & \cdots & \mathbf{S}_{N_t-1}\mathbf{F}_{2L-1} & \mathbf{S}_{N_t}\mathbf{F}_{2L-1} & \mathbf{0}_{N\times(2L-1)} & \vdots \\ \vdots & \mathbf{0}_{N\times(2L-1)} & \ddots & \ddots & \vdots & \ddots & \ddots & \mathbf{0}_{N\times(2L-1)} \\ \mathbf{0}_{N\times(2L-1)} & \cdots & \mathbf{0}_{N\times(2L-1)} & \mathbf{S}_1\mathbf{F}_{2L-1} & \mathbf{S}_2\mathbf{F}_{2L-1} & \cdots & \mathbf{S}_{N_t-1}\mathbf{F}_{2L-1} & \mathbf{S}_{N_t}\mathbf{F}_{2L-1} \end{bmatrix}, \quad (25)$$

$$\mathbf{X} = \begin{bmatrix} \mathbf{X}_1\mathbf{F}_{2L-1} & \mathbf{0}_{N\times(2L-1)} & \mathbf{0}_{N\times(2L-1)} & \mathbf{0}_{N\times(2L-1)} & \cdots & \mathbf{X}_{N_t}\mathbf{F}_{2L-1} & \mathbf{0}_{N\times(2L-1)} & \mathbf{0}_{N\times(2L-1)} & \mathbf{0}_{N\times(2L-1)} \\ \mathbf{0}_{N\times(2L-1)} & \mathbf{X}_1\mathbf{F}_{2L-1} & \mathbf{0}_{N\times(2L-1)} & \vdots & \cdots & \mathbf{0}_{N\times(2L-1)} & \mathbf{X}_{N_t}\mathbf{F}_{2L-1} & \vdots & \vdots \\ \vdots & \mathbf{0}_{N\times(2L-1)} & \ddots & \mathbf{0}_{N\times(2L-1)} & \vdots & \ddots & \mathbf{0}_{N\times(2L-1)} & \ddots & \mathbf{0}_{N\times(2L-1)} \\ \mathbf{0}_{N\times(2L-1)} & \cdots & \mathbf{0}_{N\times(2L-1)} & \mathbf{X}_1\mathbf{F}_{2L-1} & \cdots & \mathbf{0}_{N\times(2L-1)} & \cdots & \mathbf{0}_{N\times(2L-1)} & \mathbf{X}_{N_t}\mathbf{F}_{2L-1} \end{bmatrix}, \quad (26)$$

at $\mathbf{T}_1$.

According to the formulated system model in Eq. (27), it is easy found that main object of this paper is to estimate the overall channel vector $\mathbf{b}$ using the composite training signal matrix $\mathbf{D}$. With respect to Eq. (27), LS based channel estimator $\mathbf{b}_{LS}$ can be computed by

$$\mathbf{b}_{ls} = (\mathbf{D}^H\mathbf{D})^{-1}\mathbf{D}^H\bar{\mathbf{y}} = \mathbf{b} + (\mathbf{D}^H\mathbf{D})^{-1}\mathbf{D}^H\bar{\mathbf{v}}. \quad (28)$$

Since the noise variance of $\bar{\mathbf{v}}$ is given by

$$\mathrm{E}\{\bar{\mathbf{v}}^H\bar{\mathbf{v}}\} = N_0\big(\beta^2 N_r \sum_{l=0}^{L-1}\sigma_{h,l}^2 + 1\big), \quad (29)$$

then the average MSE of LS channel estimator $\mathbf{b}_{LS}$ can be given by

$$\mathrm{MSE}\{\mathbf{b}_{ls}\} = N_0\big(\beta^2 N_r \sum_{l=0}^{L-1}\sigma_{h,l}^2 + 1\big)\mathrm{Trace}\{(\mathbf{D}^H\mathbf{D})^{-1}\}. \quad (2)$$

It is well known that the training matrix $\mathbf{D}$ has $N_t(3N_t+1)(2L-1)/2$ columns that are normalized in a way such that $\|\mathbf{D}\|_F^2 = N_t(3N_t+1)(2L-1)/2$, where $\|\cdot\|_F$ denotes the Frobenius norm. Optimal training design for LS-based channel estimation method is the one that subjects to $\mathbf{D}^H\mathbf{D} = \mathbf{I}_{N_t(3N_t+1)(2L-1)/2}$. Hence, we can obtain

$$\mathrm{Trace}(\mathbf{D}^H\mathbf{D}) = \|\mathbf{D}\|_F^2 = N_t(3N_t+1)(2L-1)/2, \quad (3)$$

where $\mathrm{Trace}(\mathbf{A})$ is defined to be the sum of the elements on the main diagonal of matrix $\mathbf{A}$. According to arithmetic–harmonic means inequality, lower bound for the LS channel estimation error can be derived as

$\mathrm{MSE}\{\mathbf{b}_{ls}\}$

$$\geq \frac{N_0\big(\beta^2 N_r \sum_{l=0}^{L-1}\sigma_{h,l}^2 + 1\big)(N_t(3N_t+1)(2L-1)/2)^2}{\mathrm{Trace}\{\mathbf{D}^H\mathbf{D}\}}$$

$$= \frac{N_0\big(\beta^2 N_r \sum_{l=0}^{L-1}\sigma_{h,l}^2 + 1\big)(N_t(3N_t+1)(2L-1)/2)^2}{N_t(3N_t+1)(2L-1)/2}$$

$$= N_0\big(\beta^2 N_r \sum_{l=0}^{L-1}\sigma_{h,l}^2 + 1\big)(3N_t+1)(2L-1)/2. \quad (32)$$

From the derivation in Eq. (32), the lower bound of LS can be written as $\mathrm{MSE}\{\mathbf{b}_{LS}\} \sim \mathcal{O}(N_0, \sum_{l=0}^{L-1}\sigma_{h,l}^2, \beta, N_r, N_t, L)$. Generally, linear channel estimation methods, e.g., LS, emphasize on optimal training designing to improve estimation performance while neglect the inherent sparsity of channel.

III. SPARSE CHANNEL ESTIMATION

According to the CS [13], [14], accurate sparse channel estimation requires that training signal matrix $D$ be satisfied restricted isometry property (RIP) [15] in high probability. Hence, according to the system model in Eq. (27), optimal sparse channel estimator $\mathbf{b}_{opt}$ can be given by

$$\mathbf{b}_{opt} = \mathrm{argmin}_{\mathbf{b}}\left\{\tfrac{1}{2}\|\mathbf{D}\mathbf{b} - \bar{\mathbf{y}}\|_2^2 + \lambda\|\mathbf{b}\|_0\right\}, \quad (33)$$

where $\|\mathbf{b}\|_2$ denotes Euclidean norm which is given by $\|\mathbf{b}\|_2^2 = \sum_i |b_i|^2$; $\|\mathbf{b}\|_0$ denote zero-norm operator which counts their nonzero taps and $\lambda$ is regularization parameter which trades off the estimation error and sparseness of the channel. Assume the positions set of all channel taps of $\mathbf{b}$ is $\mathbf{\Omega}$ and its nonzero taps set is $\mathbf{\Gamma}$. The number of nonzero taps of $\mathbf{b}$ is $T$, then the lower bound of sparse channel estimator can be derived as

$$\mathrm{MSE}\{\mathbf{b}_{opt}\} = N_0\left(\beta^2 N_r \sum_{l=0}^{L-1}\sigma_{h,l}^2 + 1\right)\mathrm{Trace}\{(\mathbf{D}_\Gamma^H\mathbf{D}_\Gamma)^{-1}\}$$

$$\geq \frac{N_0(\beta^2 N_r \sum_{l=0}^{L-1}\sigma_{h,l}^2+1)T}{\mathrm{Trace}\{\mathbf{D}_\Gamma^H\mathbf{D}_\Gamma\}}$$

$$= N_0\big(\beta^2 N_r \sum_{l=0}^{L-1}\sigma_{h,l}^2 + 1\big)T, \quad (34)$$

Where $\mathrm{Trace}\{\mathbf{D}_\Gamma^H\mathbf{D}_\Gamma\} = \mathbf{I}_T$ denotes the optimal signal training for sparse channel estimation. Comparing Eq. (34) to Eq. (32), we can found that the lower bound of optimal channel estimator depends on $T$ rather than overall channel length $N_t(3N_t + 1)(2L − 1)/2$ of **b**. If we can estimate positions of nonzero taps of **b**, then sparse channel estimation performance could be improved. Since solving the optimal sparse channel estimation in Eq. (33) is NP hard problem [14]. Hence, it is necessary to develop alternative suboptimal sparse channel estimation method.

In this paper, we propose a sparse channel estimation method for MIMO-OFDM AF-TWRN and it is implemented by LASSO algorithm [10]. Given a equivalent training matrix $D$ and a received signal vector $\bar{\mathbf{y}}$, LASSO based sparse channel estimator $\mathbf{b}_{CS}$ can be obtained

$$\mathbf{b}_{cs} = \operatorname{argmin}_\mathbf{b} \left\{\frac{1}{2}\|\mathbf{D}\mathbf{b} − \bar{\mathbf{y}}\|_2^2 + \lambda\|\mathbf{b}\|_1\right\}, \quad (35)$$

where $\|\mathbf{b}\|_1$ denotes $L_1$-norm which is given by $\|\mathbf{b}\|_1 = \sum_i |b_i|$. In a practical system, accurate number of nonzero channel taps is unknown. Hence, to obtain accurate sparse channel estimation, effective training signal design is required. In accordance with the CS [13], [14], two kinds of training design methods, i.e., random Gaussian and random binary, are considered for computer simulation to evaluate our proposed method.

## IV. COMPUTER SIMULATION

In this section, we present the simulation results to evaluate sparse channel estimation method in MIMO-OFDM AF-TWRN. Here we compare the performance of the proposed estimator with LS-based channel estimator and adopt 100 independent Monte-Carlo runs for average. The number $(N_t, N_r)$ of transmitter/relay pairs are considered three cases: (2,2), (2,4) and (4,2). All of the channel vectors have same length $L = 16$ and $K = 1,2,3,4,5,6$, and its positions of nonzero channel taps are randomly generated. Training signal length of each antenna is set as $N = 32$ to ensure $N \geq 2L − 1$. Transmit power is set as $P_1 = P_2 = P$ and relay power is allocated as $P_r = 2P$. The signal to noise ratio (SNR) is defined as $10\log(P_r/\sigma_n^2)$ at relay and $10\log(P_i/\sigma_n^2), i = 1,2$ at transmitter, respectively.

Random Gaussian training is considered in Figs. 4 and 5, and random binary training is considered in Figs. 6 and 7. From the four figures, we can find that the proposed sparse channel estimator is better than LS one. In addition, the four figures show that LS channel estimator depends on channel length while LASSO one relies on nonzero number $K$ of channel. Note that the lower bound is given by ideal LS channel estimator which is known nonzero taps position of channel. In four experiments, the proposed sparse method works well on different number of nonzero taps of channel. However, for sparser channel estimation, more sparsity can be exploited. In other words, much better performance can be improved. Take the $K = 1$ for example, the proposed sparse channel estimator approach to lower bound. On the contrary, channel is approximate parse, e.g., $K = 6$, the performance advantage of the proposed method is no longer obvious. When the $K = L = 16$, then the proposed sparse channel estimator reduce to LS one. Because single channel vector between each pair of antennas is not exact sparse, it will incur much number of nonzero taps in their cascaded channel. Hence, the proposed method can works well in very sparse channel.

## V. CONCLUSION

In this paper, we proposed a sparse channel estimation method which can exploit the extra knowledge of sparse structure as for prior information and hence it can increase spectral efficient or enhance estimation performance when compared with traditional methods. Computer simulation results were showed the performance advantages of our proposed method than LS using MSE standard

ACKNOWLEDGMENT

This work was supported by the Japan Society for the Promotion of Science (JSPS) postdoctoral fellowship.

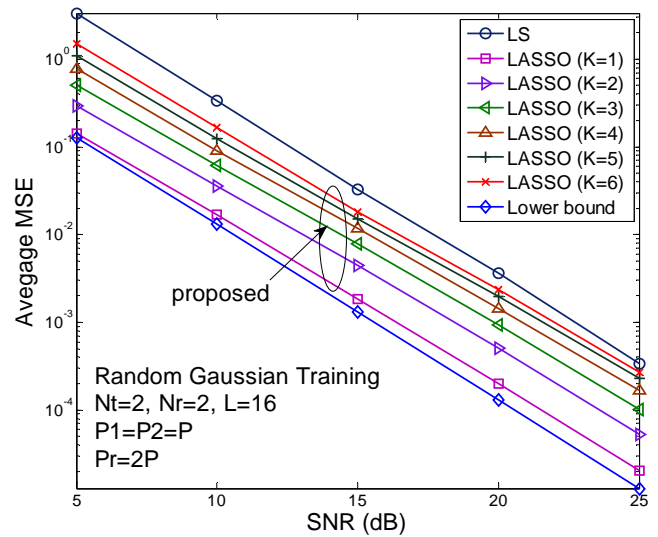

Fig. 4. Performance comparison versus SNR.

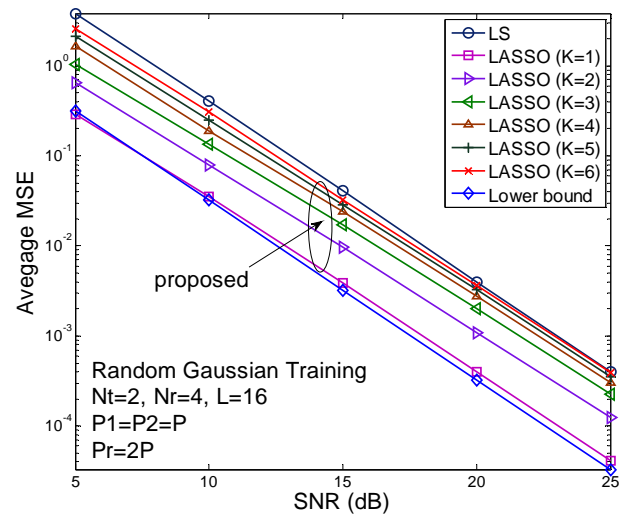

Fig. 5. Performance comparison versus SNR.

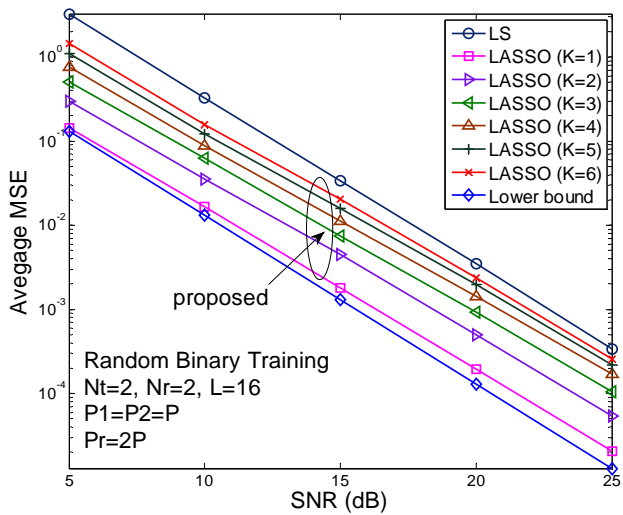

Fig. 6. Performance comparison versus SNR.

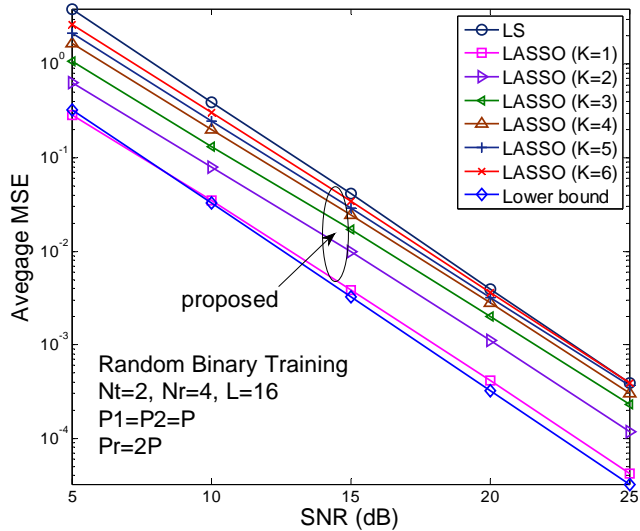

Fig. 7. Performance comparison versus SNR.